\documentclass[10pt,conference,twocolumn]{IEEEtran}

\usepackage[final]{graphicx}

\usepackage{amssymb}

\usepackage{amsmath}   
\usepackage{array}
\usepackage{epsfig}
\usepackage{balance}

\begin{document}


\title{Constellation Design for Transmission over Nonlinear Satellite Channels}

\author{\authorblockN{Farbod Kayhan}
        \authorblockA{Politecnico di Torino\\
            Dipartimento di Elettronica e Telecomunicazioni\\
           10129, Torino, Italy\\
            Email:farbod.kayhan@polito.it}\and
        \authorblockN{Guido Montorsi}
        \authorblockA{Politecnico di Torino  \\
              Dipartimento di Elettronica e Telecomunicazioni\\
              10129, Torino, Italy\\
              Email: guido.montorsi@polito.it}}


\newtheorem{Theorem}{Theorem}[section]
\newtheorem{Lemma}{Lemma}[section]
\newtheorem{Definition}{Definition}[section]
\newtheorem{Conjecture}{Conjecture}[section]
\newtheorem{Corollary}{Corollary}[section]
\newtheorem{Proposition}{Proposition}[section]

\newcommand{\C}{\mathcal{C}}
\newcommand{\SPSK}{\text{-S}\text{PSK}}
\newcommand{\SQPSK}{\text{-S}\text{QPSK}}
\newcommand{\SAPSK}{\text{-S}\text{APSK}}
\newcommand{\SkPSK}{\text{-S}^k\text{PSK}}
\newcommand{\SkQPSK}{\text{-S}^k\text{QPSK}}
\newcommand{\SkAPSK}{\text{-S}^k\text{APSK}}
\newcommand{\PSNR}{\text{PSNR}}
\newcommand{\SNR}{\text{SNR}}

\newcommand{\bin}[2]{
    \left (
        \begin{array}{@{}c@{}}
        #1 \\ #2
        \end{array}
    \right )
}

\maketitle

\begin{abstract}
In this paper we use a variation of simulated annealing algorithm for optimizing
two-dimensional constellations with 32 signals.
The main objective is to maximize the symmetric pragmatic  capacity under the peak-power constraint.
The method allows the joint optimization of constellation and binary labeling.
We also investigate the performance of the optimized constellation
over nonlinear satellite channel under additive white Gaussian noise.
We consider the performance over systems with and without pre-distorters.
In both cases the optimized constellations perform considerably better
than the conventional Amplitude Phase Shift Keying (APSK)
modulations, used in the current digital video broadcasting
standard (DVB-S2) on satellite channels. Based on our optimized
constellations, we also propose a new labeling for the 4+12+16-APSK
constellation of the DVB-S2 standard which is Gray over all rings.
\end{abstract}

\IEEEpeerreviewmaketitle

\section{Introduction}
\label{sec:INTR}
This paper investigates the constellation design for transmission over satellite channels.
Our main goal is to show that significant improvement
with respect to the Amplitude Phase Shift Keying (APSK) modulations, which
are currently used as the digital video broadcast (DVB-S2) standard, can be
obtained by optimizing the constellation set.

In this paper we focus only on the peak power (PP) constraint as
it becomes more relevant in applications with saturating
high power amplifiers as a part of the communication system.

The APSK modulation has received a considerable attention for transmission
over satellite channels. In particular, the radii and phases of each
concentric circle of APSK constellations have been optimized
in \cite{Gaudenzi} as a function of signal to noise ratio (SNR) by maximizing the capacity.
These optimized APSK constellations are accepted
as DVB-S2 standard. One of the main reasons to focus on APSK structures is
that under the PP constraint the capacity achieving distribution in the large limit
of signals is proved to be discrete in amplitude (with finite number of
mass points) and has a uniformly distributed phase in $[0,2\pi)$
\cite{Shamai_Peak}. Even though for a finite constellation the
optimal distribution is not found in general case, some previous
studies indicate that APSK modulations perform very close to
the \emph{symmetric capacity} \cite{Joint_GB2010}, \cite{PPcapacity_Tanaka}.

However, many practical applications are based on the so called
{\em pragmatic} approach: at the transmitter the
encoder is separated from the modulator using a bit interleaver
and, at the receiver, no iteration between the binary decoder and
the detector is allowed. This approach implies a loss in the capacity
that can be very small if a proper labeling of bits to constellation
points is chosen. The relevant objective function in this case
becomes the symmetric {\em pragmatic} capacity which depends on both  the constellation
set and the labeling \cite{Ungerboeck82,ZehaviPragmatic,BICM,Pragmatic_1}.
At high enough SNR, the binary reflected Gray coding has been
proved to minimize the  bit error probability for several modulation
schemes \cite{Agrell_BRGC,GrayCoding}.

Recently a joint signal-labeling optimization scheme
has been proposed by the authors \cite{Joint_GB2010} for
designing constellations which maximize the  symmetric pragmatic capacity under PP constraint
as a function of signal to noise ratio (SNR) using a simulated annealing (SA) algorithm.
However the simulated annealing algorithm becomes very slow for constellations
with 32 signals and is practically infeasible for higher order constellations.
This is mainly due to two facts. First of all, the complexity of SA algorithm grows
quadratic with respect to the number of constellations points. Secondly,
one usually needs to use a slower cooling scheduling in order to reach a \emph{good} local
maximum by increasing the constellation's cardinality.
In order to speed up the SA algorithm, we introduce a symmetry condition over the constellation
points.

We compare the performance of our optimized constellations with
32-APSK modulation used in DVB-S2 standard. We consider both \emph{static} and \emph{dynamic}
pre-distortion techniques in order to reduce the
warping effects and/or inter-symbol interference (ISI) of the non-linear
channel. Our optimized constellation outperforms the DVB-S2 current standard
up to 0.5 dB. Inspired by our optimized constellations, we also propose a new labeling
for the 4+12+16-APSK constellation which is Gray over all rings. This new mapping
results in a gain of approximately 0.15 dB with respect to the current DVB-S2
labeling for all system scenarios.

The rest of this paper is organized as follows. Section
\ref{sec:system}  describe the considered realistic satellite non linear system. In section \ref{sec:statement} we describe the main notations and formulate the optimization problem by introducing some simplifications. A short description of
the optimization algorithm and its improved version with the symmetry condition is provided in section \ref{sec:SA}.
The optimization results are presented in section \ref{sec:Results}.
We also provide a new Gray labeling for 4+12+16-APSK.
Finally, in section \ref{sec:simul} we compare the performance of
our optimized constellation with 32-APSK constellation on the realistic scenario described in Section~\ref{sec:system}.


\section{System Model}\label{sec:system}
\begin{figure}[tbh]
    \begin{centering}
        \includegraphics[angle=-90,width=0.45\textwidth]{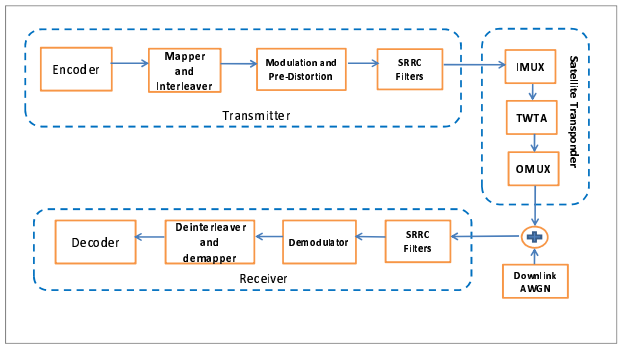}
        \caption {\label{Fig:system_model} Satellite communication system model.}
    \end{centering}
    \end{figure}

In Figure~\ref{Fig:system_model} we show the system model considered
in this paper. At the transmitter side, the information bits are passed to the
channel encoder. The codewords then are interleaved and mapped into the modulation
signals. After passing through a squared root raised cosine (SRRC) filter, the signal
is sent to the satellite transponder. In this model we ignore the uplink noise.
The transponder model is composed of an input demultiplexer (IMUX) filter,
a travelling wave tube amplifier (TWTA) and an output multiplexer (OMUX) filter \cite{DVB-S2_standard}.
The downlink noise is modelled as the complex AWGN (CAWGN). At the receiver, the signal is first
passed through the SRRC filter. Note that due to the non-linear characteristics
of the channel, this filter is no longer matched, and therefore ISI is introduced.
The decoder receives the soft estimated bits from demapper/deinterleaver and
calculates the a posteriori probabilities without any further interaction with
demodulator and demapper.

\subsection{High Power Amplifiers and Pre-Distortion Techniques}
\label{subsec:HPA}
\begin{figure}[htb]
    \begin{centering}
        \includegraphics[angle=-90,width=0.45\textwidth]{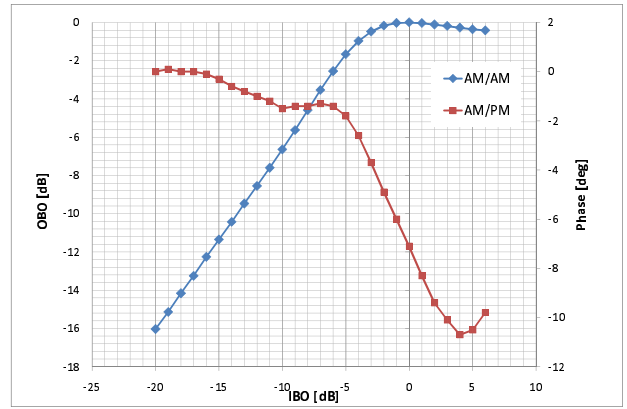}
        \caption {\label{Fig:AM/AM} AM/AM and AM/PM characteristics of DVB-S2 non linear TWTA.}
    \end{centering}
    \end{figure}

Nonlinear characteristics of the satellite channel largely depend
on the HPA used at the satellite transponder
and operating close to the saturation point. The HPA nonlinearity
changes both the amplitude and relative positions of the constellation points. These changes
are described with AM/AM and AM/PM curves depicted in figure~\ref{Fig:AM/AM}.

In some systems, the nonlinearity distortions are reduced by using a pre distorter at the
transmitter \cite{Newtec_2009}, \cite{Gaudenzi} and \cite{Karam_Sari}.
We distinguish three main categories of pre distortion techniques as follows:\\
{\bf Static symbol level predistortion}: This technique modifies
    the location of transmitted constellation  points on the complex
    plane with respect to their nominal positions in order to
    compensate the effects of the nonlinearity. The static  symbol level
    pre distorter does not modify the spectrum of transmitted signal.\\
{\bf Dynamic symbol level predistortion}: Static symbol level pre-distortion
    does not consider the impact of memory introduced by the filters in the
    system and correspondingly  neglects the ISI introduced by neighboring
    symbols.  In order to combat the ISI, one may
    takes into account also the effect of some previous and future symbols. The dynamic
    pre-distorter algorithms consider also the memory of the channel in order to compensate
    for the clustering phenomenon. Several dynamic pre-distorters have been proposed
    in the literature \cite{Gaudenzi} and \cite{CaDeGi04}.\\
{\bf Static sample level predistortion}: This technique  pre distorts
    the signal \emph{after} the transmitter shaping filter, working at
    several samples per symbol. This technique  is  essentially equivalent
    to an HPA  linearizer and can be applied only if the predistorter acts
    immediately before the HPA.
    In the following we refer to the \emph{Soft Limiter}
    or \emph{Ideal} amplifier  model, which corresponds to the use of an
    ideal sample level pre-distorter. In this case, the AM/AM curve is
    linear up to the saturating point and remains constant afterwards.

APSK constellations are more robust to the distortions caused
by nonlinear amplifier with respect to Quadrature
Amplitude Modulation (QAM).
In general the power efficiency of APSK modulation schemes can be
improved by applying pre-distortion on the transmitted signals.
Despite the hardware complexity impact,  pre-distortion techniques 
have been already adopted in systems implementing standards 
including APSK constellation such as the DVB-S2 \cite{Newtec_2009}.

\section{Statement of the optimization problem}\label{sec:statement}
We consider a complex constellation  $\chi$ with $M=2^m$ elements. The elements of
$\chi$ are referred as constellation points or simply signals.
The Euclidean distance between two points in the complex plane is denoted by
$d(.,.)$, and  $d_H(.,.)$ is used to denote the Hamming distance between two
binary sequences.

The signals are associated to the bits at the input of the
modulator through the one-to-one labeling $\mu: \chi
\rightarrow \{ 0,1 \}^m$. In particular, for any given
signal $x$,  $\mu^{i}(x)$ is the value of the $i^{th}$ bit of
the label associated to it.
A labeling for $\chi$ is called a Gray mapping if for
any two signal $x_i,x_j \in \chi$ we have
$d_H(\mu(x_i),\mu(x_j)) = 1$ if  $ d(x_i,x_j) \leq
d(x_i,x_k)$, for all $x_k \in \chi$.

Even though the separation of detection and decoding described in  section~\ref{sec:system} is in general
suboptimal, it is widespread in communications applications
due to the complexity reduction at the receiver
\cite{ZehaviPragmatic,BICM,Pragmatic_1}. The loss
in terms of channel capacity compared to optimal joint detection
and decoding, not only depends on the constellation,
but also on the labeling. In general, non-Gray mappings induce a
higher loss of capacity at high SNR's.
For a given constellation $\chi$ and labeling $\mu$, the
symmetric pragmatic capacity  of the channel is defined as
\begin{equation}
\label{eq:pragcapa}
\C(\chi,\mu) = \sum_{i=1}^{m} I(\mu^{i}(X);Y),
\end{equation}
where $\mu^{i}(X)$ is the random variable indicating the
$i^{th}$ bit associated to the transmitted signal $X$, $Y$ is
the received symbol, and $I(.;.)$ denotes the mutual
information function.
Notice that symmetric pragmatic capacity assumes uniform probability of the input symbols so that it can be equivalently named symmetric pragmatic mutual information.

\subsection{Assumptions for the Optimization Problem}
\label{subsec:assumptions}
In order to simplify the optimization problem  we make the following assumptions to model the system described in Section~\ref{sec:system}.
First of all we ignore the effect of filtering. The channel is  modelled
as $y_k = f(x_k)+n_k$
where $x_k$ is the transmitted constellation  point, $n_k$ is
the additive Gaussian noise and $f$ represents the memoryless effect of the
non linearity (AM/AM and AM/PM curves). Moreover, we assume that
the non-linearity is the \emph{Soft Limiter},
corresponding to the assumption of using  an ideal sample level predistorter:
\[
|f(x)|=\min(|x|,P_{\rm sat}=1), \;\;\;\arg(f(x))=\arg(x),
\]
where $P_{\rm sat}$ is the saturated power of the amplifier.
With this assumptions the peak-power limit imposed
over the constellation becomes equivalent to the constraint on $P_{\rm sat}$.
In section~\ref{sec:simul} we will validate the above assumptions by testing
the resulting optimized constellations in realistic
scenarios.

\section{Simulated Annealing Algorithm for Joint Signal/Labeling Optimization} \label{sec:SA}
In \cite{Joint_GB2010} a simulated annealing (SA) algorithm has
been used for maximizing $\C(\chi,\mu)$.
SA, under some conditions on the cooling schedule, guarantees convergence to the global
optimum even in non convex problems and thus it is preferable to other local optimization
techniques like the gradient algorithm.
Here we briefly review the SA algorithm presented in \cite{Joint_GB2010}.
For some other applications of SA algorithm in coding theory we refer the readers
to \cite{Salehi_TCM} and the references therein.

At any time step $t$, we randomly choose a constellation point
and move it into a random new position inside the unitary
circle. If the capacity increases by this change, we accept
the new constellation. On the other hand, if the capacity
decreases, the new constellation is accepted only with
probability $e^{-|\Delta C_t|/T_t }$, where $\Delta C_t$ is
the difference between the capacity of constellations at
times $t$ and $t-1$, and $T_t$ is the ``temperature'' at time
$t$, a parameter of the SA algorithm. The initial and final temperatures
are fixed at the beginning and the cooling parameter, $0<\alpha<1$, indicates
the cooling scheduling, i.e.,  $T_{t+1} = \alpha  T_t$.

The main adaptation needed to speed up the SA algorithm for
maximizing the capacity is to define a maximum
displacement length as a function of time. For details on such adaptation we refer
the readers to \cite{Joint_GB2010} and the references within.
Note that the initialization step distributes uniformly the $M$
signals in the unitary circle.
When the cost function is the symmetric pragmatic capacity, at the beginning of the algorithm an arbitrary labeling
is also assigned to each point and never changed afterwards, so that constellation
points "move" together with their labels during the optimization.

The output of this algorithm -with the chosen parameters-
is independent of the initial conditions for constellations with up to 16
signals, suggesting that the algorithm is optimal for such small constellations.
However for larger constellations the SA algorithm becomes very slow, and practically not applicable. Furthermore
its output starts depending on the initial conditions.

\subsection{A Simplification of Simulated Annealing with Symmetry Condition}
\label{subsec:SASymm}
In order to reduce the complexity of the SA algorithm, we introduce a symmetry condition over the constellation
points. In particular we suppose that the constellation is symmetric with respect to reflections on both horizontal and
vertical axis. To adapt the algorithm with this symmetry condition we confine the optimization
to the first quadrant of the unitary circle, i.e., we suppose that exactly one forth of points
have positive coordinates and we assume that the remaining points are obtained by reflection with respect to the two
axis. We start the algorithm by random positioning of $M/4$ signals in the first quadrant.
When the objective function is the symmetric pragmatic capacity we also  assign a random labeling of length $m-2$.
The labeling associated to reflected points is the same as the corresponding signal in first quadrant, with two additional bits selecting the quadrant. Note that if two points have Hamming distance $d$ in the first
quadrant, the corresponding reflected points also have the Hamming distance $d$ in all
other three quadrants.

The optimization technique is similar to the case without symmetry condition. At each
step of the SA we choose randomly a point from the first quadrant and move it into a new random
position inside this quadrant. Note that this change cause the simultaneous change of the three points
in the other quadrants due to the symmetry condition. In other words, at each step of SA, four points  are
simultaneously moved  inside the unitary circle.


With the addition of the symmetry condition the time needed to converge to a \emph{good}
local maximum reduces from several days to only several minutes.
Constellations obtained with this additional constraint  have the same
capacity  as those obtained without it and perform exactly the same over the non-linear satellite channel.
Furthermore, the algorithm becomes independent of the initial conditions (over ten runs). This strongly suggests
that the obtained constellations are optimal under the given conditions.

\section{Optimization Results}
\label{sec:Results}
In this section we present the constellations optimized by the
SA algorithm with symmetry condition.
As we have mentioned, we are interested in optimizing the constellations
under the peak-power constraint and therefore in what follows we always
fix the maximum power to one ($|x|^2\leq 1$). More over, instead of the signal to noise ratio,
we consider the peak power to noise ratio (PSNR) as a measure for comparing our constellations.
Note that PSNR is the ratio between the {\em peak} power of the constellation and the noise
power spectral density
$$  \PSNR \triangleq \frac{1}{N_0}\geq \SNR \triangleq
\frac{1}{M}\sum_{x\in\chi} \frac{|x|^2}{N_0}.
$$
Under the assumptions of Section~\ref{subsec:assumptions} the PSNR coincides with the ratio $P_{sat}/N_0$ so that
$$
\SNR = \frac{E_s}{N_0} = \frac{P_{\rm sat}}{N_0}\cdot\frac{E_s}{P_{\rm sat}} = \PSNR - OBO {\rm [dB]}.
$$

\begin{figure}[tbh]
\begin{centering}
    \includegraphics[angle=-90,width=0.48\textwidth]{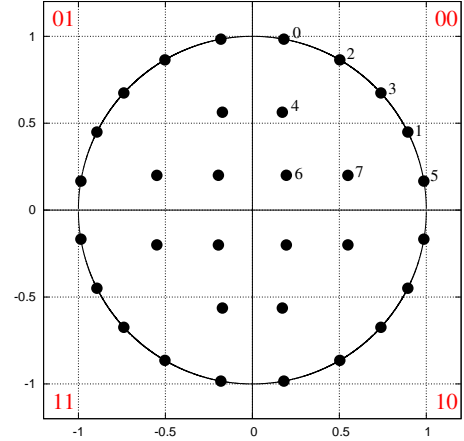}
    \caption {\label{Fig:constell_32_1} Optimized constellation at PSNR = 15 with
    the symmetry condition. }
\end{centering}
\end{figure}

Here we present two constellations optimized for two different values of PSNR. Figure~\ref{Fig:constell_32_1}
shows the constellation obtained for $\PSNR = 15$. The mapping is presented only in the first
quadrant by numbers form zero to seven. The actual mapping of a given signal in this quadrant
is a sequence of 5 bits, of which the first two bits are zero and the last three
bits correspond to the integer associated to it.
For example the point with label
$``7"$ has the mapping $``00111"$.
The mapping for signals in other quadrants are obtained by
symmetry and change of the first two bits selecting the quadrant.
Notice that the constellation in  ~\ref{Fig:constell_32_1} is by no means optimal in the classical "minimum distance" sense. At moderate and low SNR indeed the minimum distance is not dominant in the expression of the mutual information.

\begin{figure}[tbh]
\begin{centering}
    \includegraphics[angle=-90,width=0.48\textwidth]{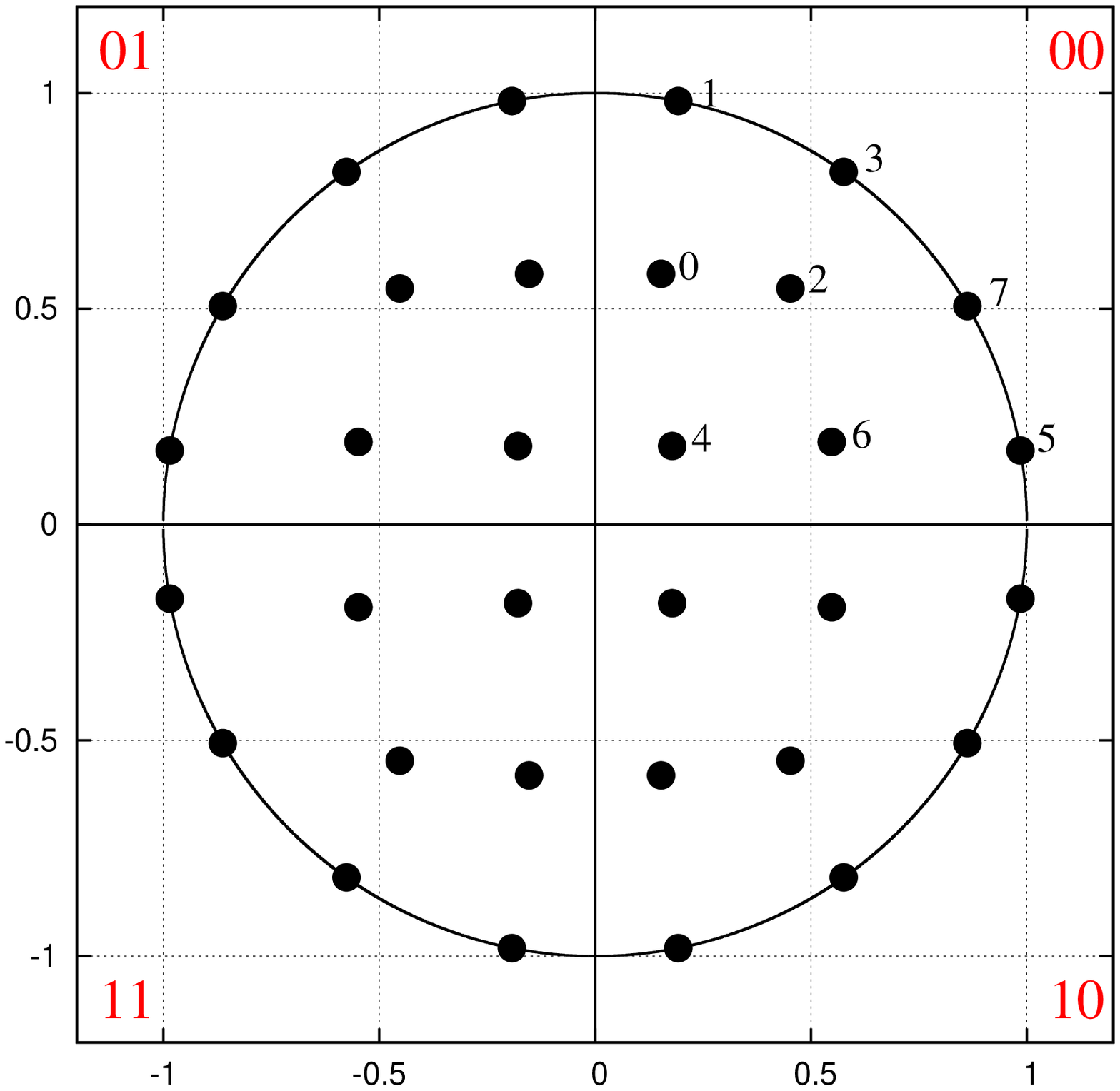}
    \caption {\label{Fig:constell_32_2} Optimized constellation at PSNR = 18 with
    the symmetry condition.}
\end{centering}
\end{figure}

In Figure~\ref{Fig:constell_32_2} we show the constellation obtained for $\PSNR = 18$
This constellation is optimized for a wide range of high PSNR values and will be used in the simulations reported in Section~\ref{sec:simul}.

\begin{figure}[tbh]
\begin{centering}
    \includegraphics[angle=-90,width=0.46\textwidth]{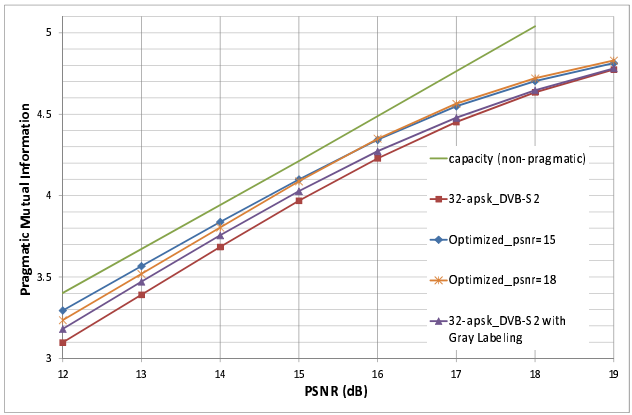}
    \caption {\label{Fig:capacity_icc} Symmetric pragmatic mutual information for the optimized
    constellations as a function of PSNR. A gain of approximately 0.5 dB is obtained
    with respect to the 32-APSK modulation scheme used currently in DVB-S2 standard.}
\end{centering}
\end{figure}

In Figure~\ref{Fig:capacity_icc} we plot the capacity curves of the two optimized constellations
as a function of PSNR and compare them with the 4+12+16-APSK modulation used currently as
the DVB-S2 standard. As it can be seen, a gain of approximately 0.5 dB is obtained by
our optimized constellations. In Section~\ref{sec:simul} we will investigate whether this gain is
preserved also in more realistic scenarios including the presence of the HPA amplifier.

The main question which arises by adding the symmetry condition over
constellation space is its impact
on the optimality of obtained constellations. We conjecture that the optimal
constellations may not satisfy the symmetry condition. The reason is mainly
because the symmetry condition
implies that the number of signals on the outer ring be divisible by four. This condition
may not need to be true in general case for an optimized constellation (see also \cite{Joint_GB2010}).

Note that one may change the ordering of the bits of a given mapping without effecting the
symmetric pragmatic capacity in equation \ref{eq:pragcapa}. When the interleaver and coding is random the ordering of
the bits of a given mapping is not an issue. However,
in DVB-S2 standard a structured row-column interleaver followed by a particular LDPC code is used and therefore it is important
to optimize also the placement of the bits \cite{DVB-S2_standard}.
This optimization can be done separately after
the joint signal/labeling optimization.

\subsection{Gray Mapping for 4+12+16-APSK}
\label{subsec:newlabeling}
\begin{figure}[tbh]
    \begin{centering}
        \includegraphics[angle=-90,width=0.48\textwidth]{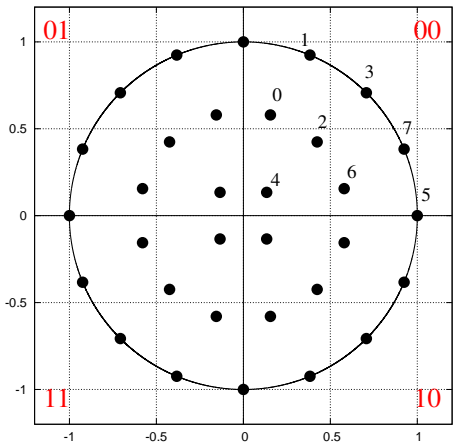}
        \caption {\label{Fig:apsk_newlabeling} New labeling for 4+12+16-APSK constellation, inspired
        by the mapping of the optimized constellation in figure \ref{Fig:constell_32_2}. Note that
        this mapping is Gray over all rings.}
    \end{centering}
    \end{figure}

Even though the constellation in Figure~\ref{Fig:constell_32_2} has not APSK structure, its
mapping can be adopted by 4+12+16-APSK constellation. Unlike the current mapping of the
DVB-S2 standard, this mapping is Gray over all rings. This new mapping for 32-APSK constellation
is shown in figure \ref{Fig:apsk_newlabeling}.
Note that we report the results obtained for this new mapping using the
4+12+16-APSK constellations with optimized radii and phases in \cite{DVB-S2_standard} and
therefore only the mapping is different from the standard.

In Figure~\ref{Fig:capacity_icc} we also plot the symmetric pragmatic capacity curve for
the 4+12+16-APSK constellation with the Gray labeling of figure \ref{Fig:apsk_newlabeling}.
This constellation has a larger pragmatic capacity than the DVB-S2 standard constellation
as a function of both SNR and PSNR. Therefore, we expect that
this constellation performs better than the current standard in all cases. This is confirmed
by our simulations over both linear and non-linear channels showing a gain of approximately
0.15 dB in all system scenarios. As we will see in the next section, the optimized constellation
performs even better than the 32-APSK with Gray labeling, suggesting that the gain obtained
by the optimization is not only due to a better mapping but also the shape of the constellation.

\section{Simulation Results}
\label{sec:simul}
In this section we investigate the performance of our optimized constellations over
a realistic system model. Systems with and without pre-distortion techniques have been
considered. We compare our results with the DVB-S2 standard 32-APSK constellation.
In all simulations we have used a LDPC code of rate $R=3/4$ and length $N=64800$. The roll-off
factor of the SRRC filter is fixed to $\alpha = 0.2$ and the normalized transponder bandwidth
(ratio between symbol rate and 3 dB bandwidth of transponder) is $\phi = 0.8$.
We use the optimized constellation in figure \ref{Fig:constell_32_2} in
all the simulations and refer to it as the \emph{optimized constellation}.
In all cases the optimal Input Back-Off (IBO) values are found preliminary and are used
for simulating the bit error rates.

In our simulation results we choose the PSNR as the measure of the quality of the channel
instead of conventional SNR. Indeed, the choice of SNR can be misleading when
comparing performance for different IBO as it only captures the loss of performance due to
the intersymbol interference (ISI) introduced by the non linearity, but it neglects the power loss
at the receiver
induced by the transponder Output Back-Off (OBO).
If one is interested in finding the corresponding average SNR, it should subtract the value
of OBO for each simulated point. Notice that since the IBO is optimized for each given PSNR
the correspondence between the two quantities is not a simple offset.

\subsection{Performance without Pre-Distortion Techniques}
\label{subsec:nopredistortion} 
\begin{figure}[tbh]
\begin{centering}
    \includegraphics[angle=-90,width=0.47\textwidth]{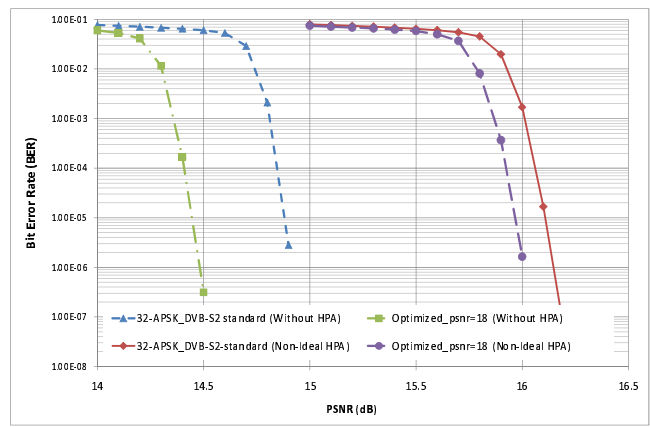}
    \caption {\label{Fig:R34_nopredistorter} Comparison between the optimized
    constellation and the 32-APSK modulation of DVB-S2 standard in the presence
    of non-ideal HPA with no pre-distortion. Also the simulation results for a system with
    no HPA is presented.}
\end{centering}
\end{figure}

We first investigate the performance of our optimized
constellation over a system where no pre-distortion technique is present. In figure
\ref{Fig:R34_nopredistorter} we compare the BER of the optimized
constellation with 32-APSK for the two different systems:
          \begin{enumerate}
            \item A system without the
          satellite transponder in the communication chain, where the channel is linear with AWGN
          noise;
            \item A system with non-ideal HPA in the communication chain but with no
          pre-distortion technique available at the transmitter.
          \end{enumerate}
As it was expected, the gain observed
in figure \ref{Fig:capacity_icc} is maintained for the system with no HPA, however this gain
is reduced into approximately 0.2 dB for the system with non-ideal HPA. As we will see
in the next subsection, this gain can be increased by applying some pre-distortion techniques.

\subsection{Performance with Pre-Distortion Techniques}
\label{withpredistortion} In this section we simulate the performance of
our optimized constellation over systems where a pre-distortion technique is
available at the transmitter. All three pre-distortion techniques discussed in
subsection \ref{subsec:HPA} are considered. For details on the implementation of
static and dynamic pre-distortion algorithms we refer the readers to \cite{CaDeGi04}.
For dynamic pre-distortion we have considered $L= 3$ symbols simultaneously for reducing
the ISI introduced by the non-linear HPA.

\begin{figure}[tbh]
\begin{centering}
    \includegraphics[angle=-90,width=0.47\textwidth]{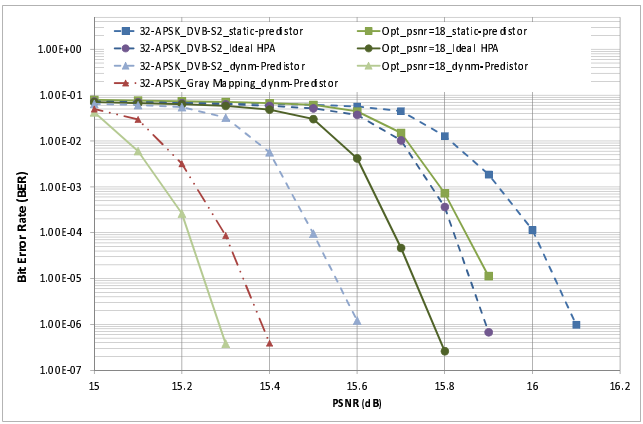}
    \caption {\label{Fig:R34_predistorter}  Comparison between the optimized
    constellation and the DVB-S2 standard 32-APSK modulation scheme in a system with
    pre-distortion. Static, dynamic and ideal pre-distortion techniques have
    been considered.}
\end{centering}
\end{figure}

In figure \ref{Fig:R34_predistorter} we present the BER for both the optimized constellation and
the 32-APSK for all discussed pre-distortion techniques. The results show that the
gain with respect to the non-ideal HPA is increased only when a dynamic pre-distorter is used,
implying that the optimized constellations suffer more from the ISI.
In this case the gain with respect to the DVB-S2 standard is slightly more than 0.3 dB. The gain which
can be obtained by using the Gray labeling for the 32-APSK constellation instead of the DVB-S2 standard
labeling is around 0.15-0.2 dB in all cases. Notice that dynamic predistortion can outperform the ``ideal'' HPA. This is due to the fact that dynamic predistortion also reduce ISI while the ideal HPA is a memoryless device that cannot reduce ISI.

\section{Conclusions}
\label{sec:conclusions}
\balance
In this paper we have proposed a simulated annealing algorithm to optimize the
constellation space under the peak-power constraint for constellations with 32
signals. We have introduced
a symmetry condition which speeds up simulated annealing algorithm, allowing to
optimize the constellations with up to 64 signals.
We have compared the performance of our optimized constellations
with 32-APSK modulations used in DVB-S2 standard in a realistic scenario.
Systems with both static
and dynamic pre-distorters have been considered. Depending on system characteristics
a gain of 0.2 to 0.5 dB can be obtained with respect to the current
DVB-S2 standard by using the optimized constellations. We have also proposed a Gray labeling for
4+12+16-APSK constellation. A gain of approximately 0.15 dB with respect to the current
labeling of DVB-S2 standard can be obtained with this new mapping.

\section*{Appendix I: Gray Mappings for APSK Constellations}
\label{App:Gray} In the large SNR region the capacity of a constellation
mainly depends on the minimum distance \cite{Foschini_2D}. In such scenarios, given a 
constellation it is desirable to find a Gray mapping for the constellation.
For a given APSK constellation the problem of finding a Gray mapping is not 
solved in general case. However it is easy to find a mapping which 
is Gray over each ring of the APSK constellation if all rings contain an even 
number of signals. If all points are nearer to a point lying on the same ring
than the points over other rings then having a Gray mapping over all rings is desirable 
in the limit of large SNR (PSNR). 

\begin{Lemma}\label{lem:Gray_map}
An APSK constellation with $2^m$ signals admits a labelling which is Gray over all rings
if and only if each ring contains an even number of points.
\end{Lemma}

\emph{Proof:} It is easy to show that no Gray mapping exists for an odd number of points over 
a ring and therefore it is necessary to have an even number of points over each 
ring of a given APSK constellation. To proof the other side of the lemma, we explicitly 
construction a mapping which is Gray for any given integer partition of $2^m$ . Let $2^m = n_1 + n_2 + ... + n_p$
be a given partition of $2^m $ where $p$ is arbitrary integer and $n_i$ is even. First we list in a column all the sequences 
with $m-1$ bits ($2^{(m-1)}$ binary numbers) such that each sequence has Hamming distance one from 
its neighbours. Then partition this list into $p$ subsets each having $n_i/2$ elements for 
$i=1...p$. Duplicate the sequences of each subset by adding a bit (both zero and one) at the 
beginning of each existing element. Now we have $p$ subsets each having exactly $n_i$ elements.
The sequences in each subset can be then arranged over a circle in such a way that the 
induced mapping is Gray. $\square$

The proof can be understood easily by looking at a simple example. Suppose that we 
want to find a Gray mapping a 4+6+8+14-APSK constellation. Such a mapping is provided
in table \ref{tab:Gray_map}.  The sequences in each section (separated by a horizontal line) can be used 
to have a Gray mapping for the ring with the corresponding number of signals.
\begin{table}
\centering{\footnotesize
\begin{tabular}{|c c|}
\hline

0 0000 & 1 0000 \\
0 0001 & 1 0001 \\
\hline
0 0011 & 1 0011 \\
0 0010 & 1 0010 \\
0 0110 & 1 0110 \\
\hline
0 0111 & 1 0111 \\
0 0101 & 1 0101 \\
0 0100 & 1 0100 \\
0 1100 & 1 1100 \\
\hline
0 1101 & 1 1101 \\
0 1111 & 1 1111 \\
0 1110 & 1 1110 \\
0 1010 & 1 1010 \\
0 1011 & 1 1011 \\
0 1001 & 1 1001 \\
0 1000 & 1 1000 \\
\hline
\end{tabular}}
\caption{ A mapping which is Gray over each ring of a 4+6+8+14-APSK Constellation following
the construction in Lemma \ref{lem:Gray_map}. }
\label{tab:Gray_map}
\end{table}

In a realistic situation with medium SNR or PSNR, one may need to have also a good distance 
property between the rings of a given APSK constellation. In this case one may first obtain a 
Gray mapping over each rings as it was explained above and then calculate the capacity of all 
possible rotations of rings choosing the rotation with highest pragmatic capacity. 
However this method may not be optimal in general case and becomes computationally infeasible 
for large constellations. 

\section*{Appendix II: Higher Order Constellations}
\label{App:64Ary} 
\begin{figure}[tbh]
\begin{centering}
    \includegraphics[angle=-90,width=0.48\textwidth]{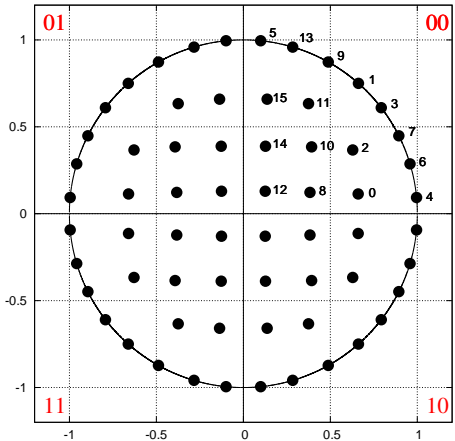}
    \caption {\label{Fig:constell_64} Optimized constellation at PSNR = 18 with
    the symmetry condition.}
\end{centering}
\end{figure}

As we have mentioned our optimization technique can be used
to optimize higher order constellations in two or more dimensions. For example 
in figure \ref{Fig:constell_64} we show the output of the algorithm for a constellation
with 64 signals optimized at $\PSNR = 18$. As it can be seen, the inner points have a cross 
32-QAM structure while 32 remaining point lie on the outer ring and the associated mapping 
is Gray.  This constellation performs slightly 
(approximately 0.15 dB) better than the 4+12+18+30-APSK constellation for a LDPC code with 
rate $R=3/4$ and length $N=64800$ over the non-linear channel with ideal amplifier.
More research on this direction is on going.

\section*{Acknowledgment}
The authors wish to thank Sergio Benedetto, Nader Alagha and Riccardo De Gaudenzi
for several discussions and useful suggestions. This work was performed under the ESA/ESTEC
contract n. $4000102300$,  ``Enhanced Digital Modem Techniques Development and Validation".
\bibliographystyle{IEEEtran}
%
%

\end{document}